\documentclass{aastex63}

\usepackage{epsfig}
\renewcommand{\arraystretch}{1.2}

\received{Month DD, 2020}
\revised{Month DD, 2020}
\accepted{Month DD, 2020}
\submitjournal{ApJ}
\shorttitle{D$_3$ Line Polarization}
\shortauthors{Heinzel et al.}

\begin{document}
\title{On the Possibilities of Detecting Helium D$_3$ Line Polarization with Metis}

\correspondingauthor{Petr Heinzel}
\email{pheinzel@asu.cas.cz}

\author[0000-0002-5778-2600]{Petr Heinzel}
\affiliation{Astronomical Institute, The Czech Academy of Sciences, 25165 Ond\v{r}ejov, Czech Republic}

\author[0000-0002-8292-2636]{Ji\v{r}i \v{S}t\v{e}p\'{a}n} 
\affil{Astronomical Institute, The Czech Academy of Sciences, 25165 Ond\v{r}ejov, Czech Republic}

\author[0000-0001-5796-5653]{Alessandro Bemporad} 
\affiliation{INAF - Turin Astrophysical Observatory, Via Osservatorio 20, 10025 Pino Torinese, TO, Italy}

\author[0000-0002-2789-816X]{Silvano Fineschi} 
\affiliation{INAF - Turin Astrophysical Observatory, Via Osservatorio 20, 10025 Pino Torinese, TO, Italy}

\author[0000-0001-8489-4037]{Sonja Jej\v{c}i\v{c}} 
\affiliation{Faculty of Education, University of Ljubljana, Kardeljeva plo\v{s}\v{c}ad 16, 1000 Ljubljana, Slovenia} 
\affiliation{Faculty of Mathematics and Physics, University of Ljubljana, Jadranska 19, 1000 Ljubljana, Slovenia}
\affiliation{Astronomical Institute, The Czech Academy of Sciences, 25165 Ond\v{r}ejov, Czech Republic}

\author[0000-0002-4638-157X]{Nicolas Labrosse}
\affiliation{SUPA, School of Physics \& Astronomy, University of Glasgow, Glasgow, G12 8QQ, UK}

\author[0000-0002-1017-7163]{Roberto Susino}
\affiliation{INAF - Turin Astrophysical Observatory, Via Osservatorio 20, 10025 Pino Torinese, TO, Italy}

\begin{abstract}
Space coronagraph Metis on board of the Solar Orbiter offers us new capabilities for
studying eruptive prominences and coronal mass ejections (CME). Its two spectral channels,
hydrogen L$\alpha$ and visible-light (VL) will provide, for the first time, co-aligned and co-temporal
images to study dynamics and plasma properties of CMEs. Moreover, with the
VL channel (580 - 640 nm) we find an exciting possibility to detect the helium D$_3$ line (587.73 nm)
and its linear polarization. The aim of this study is to predict the diagnostics potential of
this line regarding the CME thermal and magnetic structure. For a grid of
models we first compute the intensity of the D$_3$
line together with VL continuum intensity due to Thomson scattering on core electrons. We show that
the Metis VL channel will detect a mixture of both, with predominance of the helium emission at intermediate
temperatures between 30 - 50,000 K. Then we use the code HAZEL to compute
the degree of linear polarization detectable in the VL channel. This is a mixture of D$_3$ scattering
polarization and continuum polarization. The former one is lowered in the presence of a magnetic field
and the polarization axis is rotated (Hanle effect). Metis has the capability of measuring $Q/I$ and 
$U/I$ polarization degrees and we show their dependence on temperature and magnetic field. At $T$=30,000 K
we find a significant lowering of $Q/I$ which is due to strongly enhanced
D$_3$ line emission, while depolarization at 10 G amounts roughly to 10 \%.  
\end{abstract}

\section{Introduction}

Magnetic-field measurements in cool coronal structures like prominences, coronal rain or CMEs represent
a challenging problem. In case of quiescent prominences, several attempts were made to determine the
supporting magnetic fields using the spectro-polarimetry - for a review see \cite{Lopez2015}.
Prominences are low-density media and thus the scattering of the incident solar radiation determines
their emissivity. Their illumination by the solar disk is largely anisotropic which leads to
a linear polarization of the scattered radiation. The presence of a magnetic field, which is rather
weak in prominences, then causes a lowering of the polarization degree and rotation of the
polarization plane (Hanle effect, see \cite{LL04}). Typical range of the magnetic-field strength in quiescent
prominences is around 10 G, but fields as high as tens of G have also been reported \citep{Lopez2015}. 
Recently, much stronger field was detected
in post-flare loops, earlier called loop prominences and nowadays identified with a coronal rain.
However, in case of flare loops with the field strength of a few hundreds of Gauss,
the polarization is due to Zeeman effect and the weak-field approximation was used to determine the field
strength \citep{Kuridze+2019}. Another class of relatively cool coronal structures is represented
by cores of Coronal Mass Ejections (CME), where the kinetic temperatures are on the order
10$^4$ to 10$^5$ K or more (see e.g. \cite{Heinzel+2016} and \cite{Jejcic+2018}). In these structures,
however, the magnetic field was never measured. This is related to the fact that CMEs are mostly
observed from space, using coronagraphs in visible light (e.g. LASCO on board SOHO and COR1 and COR2
on board STEREO missions) or with
the UV spectrograph like UVCS on board SOHO. 
Nevertheless, there are no spectro-polarimeters attached to space coronagraphs capable of measuring the
linear polarization in spectral lines emitted by prominence-like structures including CME cores.
A giant coronagraph ASPIICS on board the ESA's Proba-3 formation-flight mission \citep{Lamy+2017} will
detect the helium D$_3$ line at 587.73 nm (vacuum wavelength), but only the integrated intensity and no polarization.
D$_3$ line polarization in prominence-like structures was studied in many cases, both theoretically
as well as observationally (see \cite{Lopez2015} and, therefore, some attempts were made
to include D$_3$ polarization measurements in the concept of ASPIICS. But due to different reasons
the final set-up will provide polarization detection only in the broad-band visible channel important for determinations
of the electron density. 

With the launch of the ESA-NASA Solar Orbiter mission, we find an exciting possibility to detect the
D$_3$ line polarization in eruptive prominences and CMEs, 
using the Metis coronagraph (for Metis description see \cite{Antonucci+2019}). 
This is because the visible-light (VL) continuum channel of Metis in the range between 580 - 640 nm contains the D$_3$ line
at its wavelength edge, still well detectable, and this channel will provide the polarization
measurements. Note that Metis has another imaging capability in the hydrogen Lyman $\alpha$ line,
which will provide important diagnostics of the CMEs and coronal plasmas. The situation with
Metis VL channel is similar to that of SOHO/LASCO-C2, where the orange VL channel also
contains the helium D$_3$ line. Quite recently, \cite{FloydLamy2019}
analyzed several CMEs detected in the orange channel of LASCO-C2 and they discuss apparent signatures 
of the D$_3$ polarization. 
On the other hand, \cite{Dolei+2014} were able to 
extract the H$\alpha$ line polarization in a CME combining STEREO-COR1 and LASCO-C2 observations and they suggested that
this could be used to determine the magnetic field in CMEs. 

Our idea is to use the Metis VL channel to detect the D$_3$ polarization in CMEs and possibly
to measure their magnetic fields. The main difficulty is that the expected line polarization
is mixed with the linear continuum polarization which is due to Thomson scattering on CME
electrons. In this paper we estimate theoretically the amount of D$_3$ line polarization under
typical CME conditions and compare it with the respective continuum component. Then we discuss
possibilities of the magnetic-field determination based on Metis observations.

\section{Models of Eruptive Prominences and CMEs}

\begin{table}
\centering{
\caption{Grid of models used for synthesis of polarized radiation in D$_3$ line and VL.}
\label{table:models}
\begin{tabular}{cccccccccc}
\hline \\[-2.8ex]
  model & $h$ & $T$ & $p$ & $D$ & $v_{\rm t}$ & $n_{\mathrm{e}}$ & $E$(D$_3$) & $\tau_0$(D$_3$) & $E$(VL)\\
            & km & K & dyn $\mathrm{cm}^{-2}$ & km & km s$^{-1}$ & $\mathrm{cm}^{-3}$ & cgs & & cgs \\
\hline \\[-2.8ex]
  1 & 800000 & 8,000 & 0.05 & 5000 & 5 & 3.83+9 & 44.2 & 8.4-4 & 90.3 \\
  2 & 800000 & 15,000 & 0.05 & 5000 & 5 & 1.15+10 & 34.9 & 5.2-4 & 271.0\\
  3 & 800000 & 30,000 & 0.05 & 5000 & 15 & 6.29+9 & 885.0 & 6.7-3 & 148.2\\
  4 & 800000 & 50,000 & 0.05 & 5000 & 15 & 3.78+9 & 176.9 & 1.2-3 & 89.1\\
  5 & 800000 & 100,000 & 0.05 & 5000 & 20 & 1.89+9 & 2.8 & 1.6-5 & 44.5\\
  6 & 800000 & 8,000 & 0.1 & 5000 & 5 & 8.35+9 & 35.8 & 6.8-4 & 196.8\\
  7 & 800000 & 15,000 & 0.1 & 5000 & 5 & 2.34+10 & 66.1& 9.5-4 & 551.5\\
  8 & 800000 & 30,000 & 0.1 & 5000 & 15 & 1.26+10 & 2244.4 & 1.6-2 & 297.0\\
  9 & 800000 & 50,000 & 0.1 & 5000 & 15 & 7.56+9 & 439.8 & 2.7-3 & 178.2\\
10 & 800000 & 100,000 & 0.1 & 5000 & 20 & 3.78+9 & 6.9 & 3.6-5 & 89.1\\
11 & 1600000 & 8,000 & 0.05 & 5000 & 5 & 1.75+9 & 7.5 & 3.3-4 & 16.5\\
12 & 1600000 & 15,000 & 0.05 & 5000 & 5 & 1.13+10 & 6.9 & 2.0-4 & 106.5\\
13 & 1600000 & 30,000 & 0.05 & 5000 & 15 & 6.29+9 & 290.4 & 3.7-3 & 59.3\\
14 & 1600000 & 50,000 & 0.05 & 5000 & 15 & 3.78+9 & 57.9 & 6.2-4 & 35.6\\
15 & 1600000 & 100,000 & 0.05 & 5000 & 20 & 1.89+9 & 0.9 & 8.2-6 & 17.8\\
16 & 1600000 & 8,000 & 0.1 & 5000 & 5 & 4.07+9 & 6.1 & 2.6-4 & 38.4\\
17 & 1600000 & 15,000 & 0.1 & 5000 & 5 & 2.31+10 & 20.3 & 5.3-4 & 217.7\\
18 & 1600000 & 30,000 & 0.1 & 5000 & 15 & 1.26+10 & 850.9 & 8.7-3 & 118.8\\
19 & 1600000 & 50,000 & 0.1 & 5000 & 15 & 7.56+9 & 170.0 & 1.4-3 & 71.3\\
20 & 1600000 & 100,000 & 0.1 & 5000 & 20 & 3.78+9 & 2.6 & 1.8-5 & 35.6\\
\hline
\end{tabular}
}
\end{table}

For the purpose of this exploratory work we use the prominence-CME models as described by
\cite{Jejcic+2018} who studied the capabilities of the narrow-band D$_3$ filter for ASPIICS. 
Since the temperatures in those models range from 8,000 K up tp 10$^5$ K, the models we
select here can represent cool erupting prominence plasma as well as hot cores of CMEs \citep{Heinzel+2016}.
The electron densities $n_{\rm e}$ are first computed with the hydrogen code and then used to synthesize the helium lines
as in 
\cite{LabrosseGouttebroze2001} and \cite{LabrosseGouttebroze2004}.
Note that we neglect here the effect of potentially large CME velocities on the electron density, an aspect to be considered
in a future modeling. Such velocities, however, do not affect the formation of the D$_3$ line because the prominence is
illuminated by a continuum radiation, i.e. no Doppler brightening effect takes place. 
D$_3$ line-center optical thickness $\tau_0$ is the input parameter for our polarized radiative-transfer modeling.
We need to know the relation
between $\tau_0$ and $n_{\rm e}$ in order to consistently evaluate the D$_3$ and VL emissions which enter the Metis
filter passband. From a grid of 90 models computed in \cite{Jejcic+2018} we selected 20 representative ones as shown
in Table~\ref{table:models}. We choose two heights above the solar surface, and namely 800 Mm with geometrical dilution factor $W$=0.058 and 1600 Mm with $W$=0.024, 
which correspond to 2.15 and 3.30 solar radii measured from the disk center, respectively, in the
range covered by Metis at its closest approach to the Sun. The dilution factor $W$
substantially decreases with height which lowers the amount of exciting radiation but simultaneously increases its
anisotropy. For the effective geometrical thickness $D$ we choose 5 Mm which, together with a low filling factor
\citep{Susino+2018}, represents plausible sizes of CME cores. The selected gas pressures give the hydrogen ionization
comparable to situations in eruptive prominences and hot CMEs. Similarly as in \cite{Jejcic+2018} we increase the 
microturbulent velocity with increasing temperature (see the case or a CME flux-rope in \cite{Jejcic+2017}).

\section{Helium D$_3$ Line Formation in Prominences and CMEs}

\begin{figure} 
\begin{center}
\includegraphics[width=18.0cm]{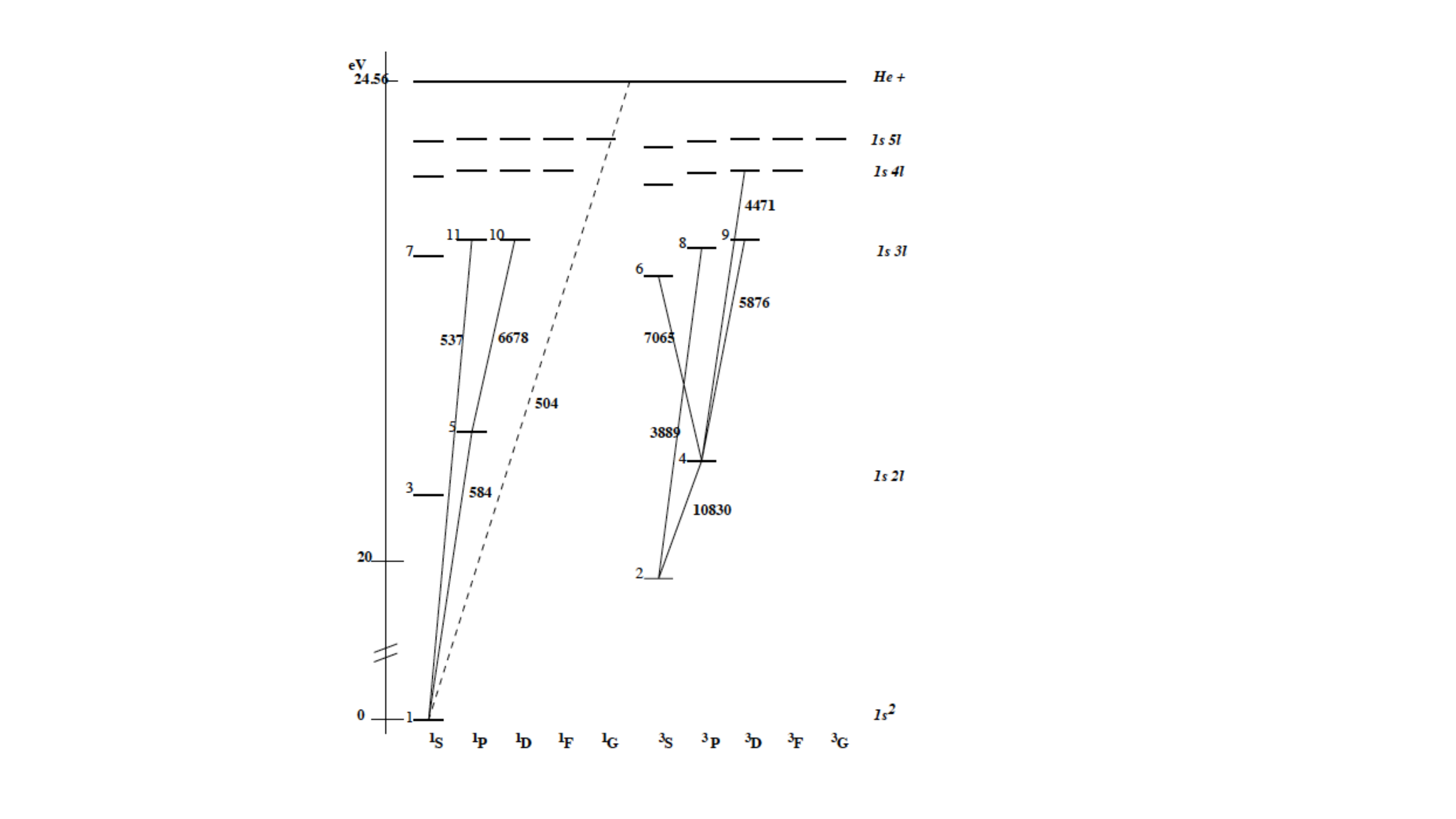} 
\end{center}
\caption{Atomic level and transition diagram for \ion{He}{1} atom. The wavelengths of line transitions
are indicated, the dashed line represents the ionization continuum from the ground state. This line 
also schematically divides the singlet and triplet states of \ion{He}{1}. Here the D$_3$ line is due to transition
between levels 4 and 9.}
\label{atom}
\end{figure}

Formation of the helium D$_3$ line under non-LTE conditions (i.e. departures from the Local Thermodynamic Equilibrium) is
a complex multilevel radiative-transfer problem. In case of solar prominences it was treated in detail by \cite{Heasley+1974}
and later on by \cite{LabrosseGouttebroze2001} and \cite{LabrosseGouttebroze2004}
(see also reviews by \cite{Labrosse+2010} and \cite{Labrosse2015}). The latter authors used a multi-level 
\ion{He}{1} model atom depicted in Figure \ref{atom} and solving the multi-ion statistical equilibrium equations they 
obtained the ionization structure, level populations and optical properties of the helium. Here we will focus only on the formation
of the D$_3$ line, which can be approximately separated into two problems: the excitation of level 9 from which the optically-thin D$_3$
emission arises, and population of the lower level 4 which determines
the optical thickness $\tau_0$ of the D$_3$ line (see Table~\ref{table:models}). 
While the first aspect can be treated as a two-level atom problem, the second one is a complex multilevel-multiion non-LTE problem.
Optical thickness $\tau_0$
and integrated intensity of D$_3$ result from the helium multilevel modeling as described in \cite{LabrosseGouttebroze2001}
(note that the helium non-LTE code uses as input the electron density previously computed with the hydrogen code).
These quantities are shown in Table~\ref{table:models}, together with the visible-light (VL) intensity integrated over the Metis VL passband
580 - 640 nm. VL emission is due to Thomson scattering on prominence or CME electrons and was computed using the 
limb-darkened incident continuum radiation from the solar disk (e.g. \cite{Cox2000}). We can see from Table~\ref{table:models} that even
using this wide-band Metis filter, the D$_3$ line intensity is not negligible in comparison to VL intensity and namely for
higher temperatures around 30 - 50,000 K  the D$_3$ intensity dominates the VL one. This means that we may expect
a non-negligible contribution of D$_3$ to total polarization signal from Metis filter. This is also consistent with the
conclusions of \cite{FloydLamy2019} who found CME signatures of the D$_3$ emission within even wider (100 nm) broad-band orange filter
of LASCO-C2 coronagraph.
First question arises how the upper level 9 of the D$_3$ transition is excited. In \cite{Jejcic+2017}
the authors suggest that much brighter D$_3$ at $T$=30,000 K can be due to stronger collisional excitations at higher temperatures.
This of course would produce a non-polarized emission, i.e. the scattering term will be negligible compared to collisional one in the
line source function. We know that collisions,  both inelastic as well as elastic, are negligible at typical prominence temperatures below
say 10,000 K, but how it will be at much higher temperatures found in CME cores ? In order to answer this critical
question, we made two calculations. First, we quantitatively compared the collisional excitation
rates to upward radiative ones (i.e. those leading to scattering). Their ratio is 
\begin{equation}
x = \frac{n_{\rm e} C_{49}}{B_{49} I_0 W} \, ,
\end{equation}
where $B_{49}$ is the Einstein coefficient for absorption, $I_0$ the disk-center continuum intensity at the D$_3$ wavelength
and $W$ the geometrical dilution factor (in this estimate we neglect the continuum limb darkening). $C_{49}$ is the collisional
rate depending on temperature. We found that for all models considered here this ratio is quite small, the radiative rates are
several orders of magnitude larger than the collisional ones, even at high temperatures. This is a good news since the line
source function in this two-level model is dominated by scattering. We also found negligible
collisional rates for transition between levels 2 and 4 which means that the level 4 is populated by radiative excitation
and thus is polarized (see the next section).

The other independent calculation shows that actually all D$_3$ line intensities from Table~\ref{table:models} result from almost identical line
source function which is dominated by scattering. For the line integrated intensity we can simply write

\begin{equation}
E = \sqrt\pi \Delta\lambda_{\rm D} S_{\lambda} \tau_0 \, ,
\end{equation}
where $\Delta\lambda_{\rm D}$ is the line Doppler width and $S$ is the line source function. Using $E$, $\Delta\lambda_{\rm D}$ and $\tau_0$ 
according to Table~\ref{table:models}, we find that resulting source function $S$ is almost identical for all models. This then means that also at higher temperatures
it must be dominated by scattering.

However, there is still a question
why D$_3$ brightness is so large at temperatures around 30,000 K and this is the other aspect of the D$_3$ line 
formation problem. Looking at Eq. 2 we see that, for a fixed $S$, $E$ varies due to changes of $\Delta\lambda_{\rm D}$ and $\tau_0$.
However, the product $\Delta\lambda_{\rm D} \times \tau_0$ is directly proportional to number density of \ion{He}{1} atoms in level 4, i.e. the
level 4 population. Since the level 4 is mainly populated by radiative excitations from level 2, i.e. the scattering in 1083 nm line (collisional excitation
is again negligible at low densities),  the temperature dependence of $\tau_0$(D$_3$) must follow that of the 2nd level population.
It is generally known that this particular triplet state is populated by recombinations (radiative and di-electronic) from
\ion{He}{2} ion. Therefore, it must depend on the ionization rate from \ion{He}{1} to \ion{He}{2}. If this population should be dependent on temperature, the
collisional ionization of \ion{He}{1} from its ground state must dominate over the radiative one. We thus computed these two rates. The 
radiative (photoionization) rate was estimated using the incident EUV ionizing radiation below 50.4 nm and the collisional ionization rates
were computed according to \cite{MihalasStone1968}. Very interestingly, around $T$=30,000 K, the collisional ionization is already dominant
over the photoionization by almost one order of magnitude while at low temperatures it is quite negligible. 
We may thus expect a significant temperature-dependent increase of \ion{He}{2} density and thus
also of population of the triplet ground state 2 due to photo-recombinations. 
However, for much higher temperatures reaching 10$^5$ K, $\tau_0$(D$_3$) will substantially
decrease due to strong ionization of helium (see Table~\ref{table:models}).  In summary, we see that $E$(D$_3$) is
varying with temperature due to \ion{He}{1} collisional ionization and subsequent photo-recombinations, but the line source function is completely
dominated by scattering. Note that a slight difference from such a source function at high temperatures is probably due to recombinations 
directly to level 4.

\section{Visible-Light and D$_3$ Scattering Polarization} 

\subsection{Visible-light component}

\begin{figure}
\begin{center}
\includegraphics[width=6.0cm]{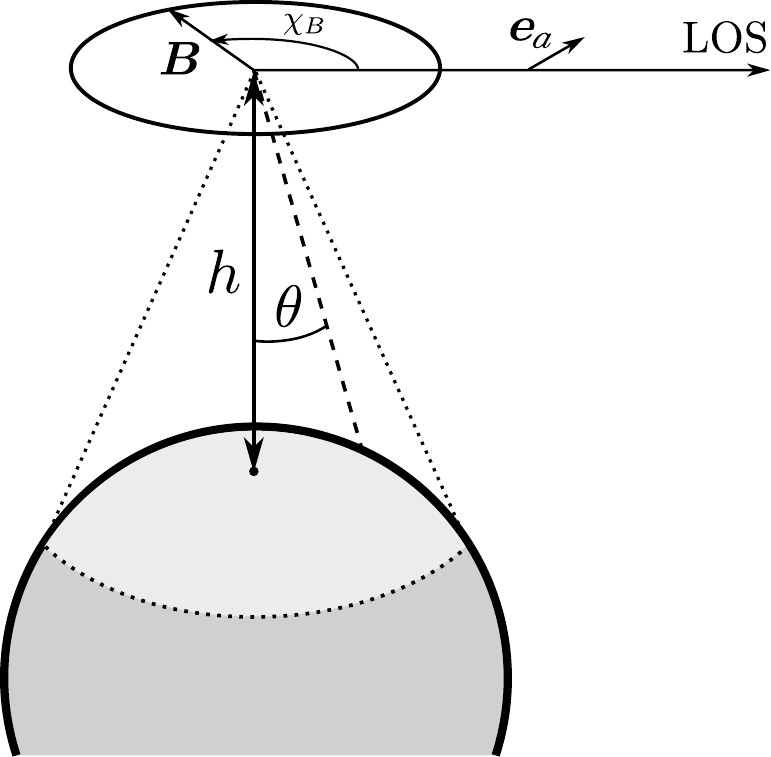}
\end{center}
\caption{Scattering geometry considered in this paper. The plasma is located at the height $h$ above the solar surface and scatters the incident disk radiation that arrives at the angle $\theta$ between the local vertical and the direction of illumination. The positive Stokes $Q$ direction (i.e., the $\vec e_a$ vector) is parallel to the nearest solar limb. The magnetic field vector $\vec B$ is perpendicular to the solar radius and deviates by an angle $\chi_B$ from the line of sight (LOS) that is chosen to be perpendicular to the local vertical direction.}
\label{fig:scattgeo}
\end{figure}

\begin{figure} 
\begin{center}
\includegraphics[width=10.0cm]{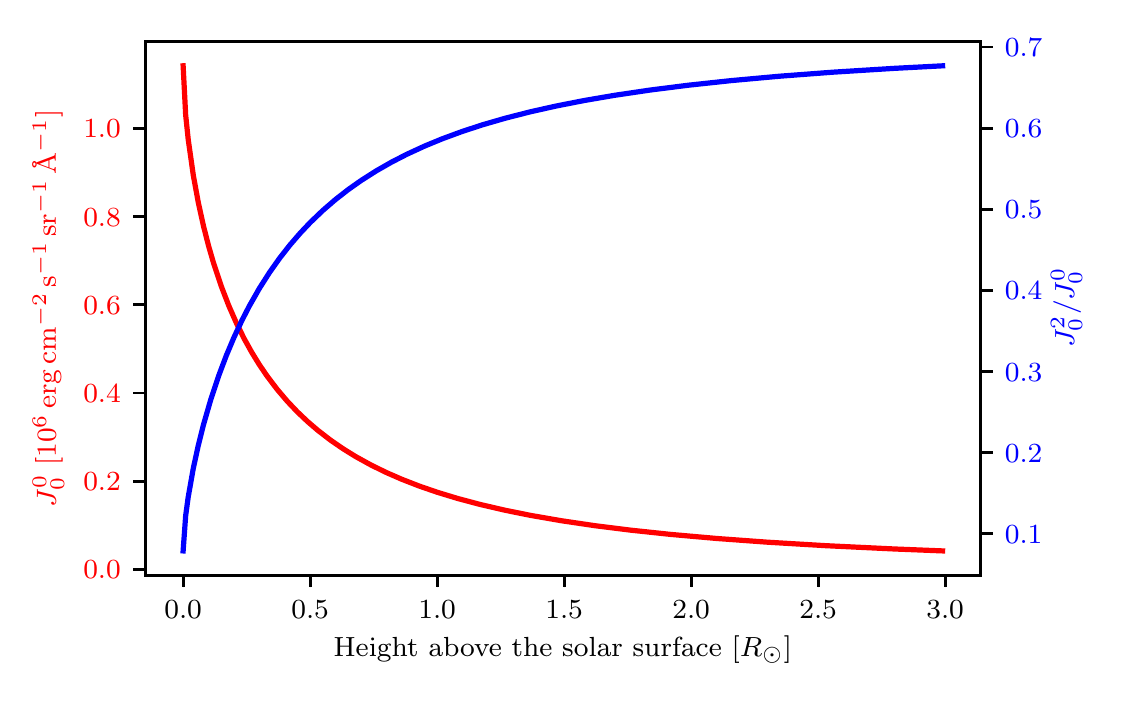} 
\end{center}
\caption{Mean intensity ($J^0_0$) and fractional anisotropy ($J^2_0/J^0_0$) as a function of height above the solar surface at the wavelength $\lambda=600$\,nm.}
\label{fig:jkq}
\end{figure}

For the sake of simplicity, we consider a simple scattering geometry shown in Figure~\ref{fig:scattgeo}. The plasma located in the plane of the sky at the height $h$ above the solar surface is illuminated by an anisotropic radiation from the underlying solar photosphere. We assume the illumination is due to an unpolarized solar continuum radiation in the whole spectral interval of the VL filter. Due to the limb darkening effect, the incindent intensity $I^{\rm inc}(\lambda,\theta)$ at wavelength $\lambda$ depends on the incident angle $\theta$. In our calculations, we use interpolated data from \cite{Cox2000} for $I^{\rm inc}(\lambda,\theta)$.

The VL continuum emission is predominantly due to the Thomson scattering. In order to calculate the intensity and linear polarization of the scattered radiation at a given height and wavelength, two components of the radiation field tensor need to be considered, namely $J^0_0$ that corresponds to the common mean radiation field intensity $J$, and $J^2_0$ that quantifies anisotropy of the incident field:
\begin{eqnarray}
J^0_0(\lambda)&=&\frac 12 \int_{-1}^1 I^{\rm inc}(\lambda,\mu) \; d\mu\,, \\
J^2_0(\lambda)&=&\frac{1}{4\sqrt 2}\int_{-1}^1 (3\mu^2-1)I^{\rm inc}(\lambda,\mu)\, d\mu\,,
\end{eqnarray}
where $\mu=-\cos\theta$ (for more details, see \cite{LL04}). In Figure~\ref{fig:jkq}, we show the height dependence of these quantities at the wavelength $\lambda=600\,{\rm nm}$. Even though these quantities depend on wavelength, this dependence is rather weak in the interval of interest (580 --- 640 nm). As it follows from the plot, the fractional anisotropy
$J^2_0/J^0_0$ rapidly increases with height and so is the fractional polarization of the emited VL radiation (see below).

The continuum optical thickness of the slab of free electrons is independent of wavelength and equal to $\tau_{\rm e}=n_{\rm e}\sigma_{\rm T}D$, where $n_{\rm e}$ is the electron number density, $\sigma_{\rm T}\approx6.65\times 10^{-25}\,{\rm cm^{-2}}$ is the Thomson scattering cross-section, and $D$ is the geometrical thickness of the slab. In the models considered in this paper, we are always in the regime of very small optical thickness, $\tau_{\rm e}\ll 1$. In that case, the scattered continuum intensity (Stokes parameter $I$) and linear polarization (Stokes parameter $Q$) in the geometrical configuration of Figure~\ref{fig:scattgeo} are equal to the respective Stokes source functions $S_I$ and $S_Q$ multiplied by the optical thickness of the medium. The source functions of the Stokes parameters can be easily derived \citep[e.g.,][]{jtb09} and the expressions for the emergent Stokes parameters read
\begin{eqnarray}
I(\lambda)&=& \tau_{\rm e} S_I(\lambda)=\tau_{\rm e} J^0_0(\lambda) \left[1-\frac{1}{2\sqrt 2}\frac{J^2_0(\lambda)}{J^0_0(\lambda)} \right]\,, \label{eq:si} \\
Q(\lambda)&=& \tau_{\rm e} S_Q(\lambda)=\tau_{\rm e} \frac{3}{2\sqrt 2} J^2_0(\lambda)\,. \label{eq:sq}
\end{eqnarray}
We note that at heights above $h\approx 0.1R_{\odot}$, the second term in Eq.~(\ref{eq:si}) is not negligible (see Figure~\ref{fig:jkq}) and since $J^2_0>0$, the emited intensity is lower than one would expect if anisotropy and polarization phenomena were neglected.

Integration of the above expressions over the Metis VL pass-band gives us the observable total emisivities of VL in intensity and linear polarization,
\begin{eqnarray}
E_I&=&\int I(\lambda) \phi(\lambda) \, d\lambda\,, \\
E_Q&=&\int Q(\lambda) \phi(\lambda) \, d\lambda\,,
\end{eqnarray}
where $\phi(\lambda)$ is a normalized spectral sensitivity of the instrument. For the sake of simplicity, we consider $\phi(\lambda)=1$
in the range between 580 - 640 nm (see \cite{Antonucci+2019}).

The linear polarization of the VL is parallel to the nearest solar limb, i.e., the Stokes parameter $U$ and $E_U=\int U(\lambda)\phi(\lambda)\,d\lambda$ are equal to zero, and it is insensitive to presence of magnetic field. In contrast to Thomson scattering, the linear polarization of the D$_3$ line is sensitive to the magnetic fields via the Hanle and Zeeman effects. If the wavelength-integrated signal is contaminated by the photons emited by the \ion{He}{1} atoms, the $E_I$, $E_Q$, and $E_U$ can, in principle, provide information on the magnetic field vector in the slab.

\subsection{Spectral line component}

\begin{figure} 
\begin{center}
\includegraphics[width=15.0cm]{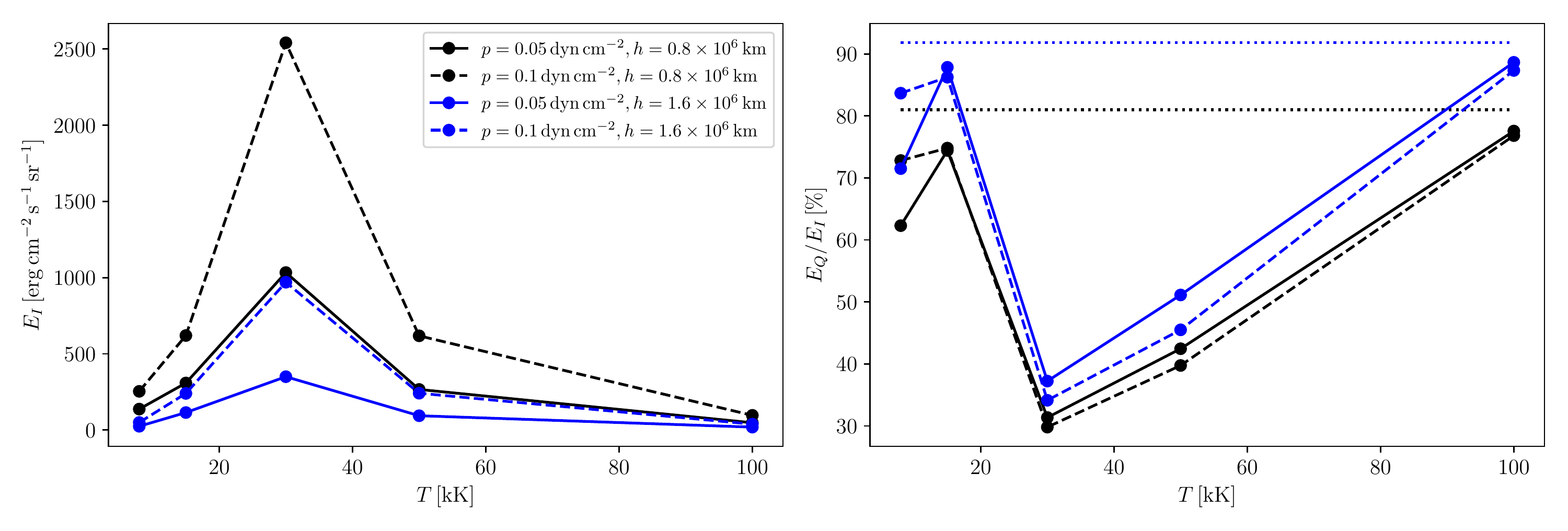} 
\end{center}
\caption{Left panel: integrated intensity $E_I$(D$_3$+VL). Right panel: integrated linear-polarization degree $E_Q$(D$_3$+VL)/$E_I$(D$_3$+VL) signal. The signals for two different heights above the solar surface and two different plasma pressures are plotted as functions of kinetic temperature. The horizontal dotted lines in the right panel show the fractional polarization of the VL, neglecting the D$_3$ contribution, at the heights $h=0.8\times 10^6$\,km (black line) and $1.6\times 10^6$\,km (blue line).}
\label{fig:iqb0}
\end{figure}

As shown by \citet{bommier77}, once the density of orthohelium is known from the non-LTE calculation, the atomic model sufficient for synthesis of the D$_3$ line intensity and polarization consists of the terms 2, 4, 6, 8, 9
in Figure~\ref{atom} with total 11 fine-structure levels. In the low-density plasma of our interest, the depolarizing collisions can be neglected \citep{1977A&A....59..223S}. Since the optical thickness of D$_3$ (and presumably of the other considered lines) is smaller than one, cf. Table~\ref{table:models}, and since the incident photospheric radiation is spectrally flat across the D$_3$ line and, to a large extent across the other relevant lines of the model atom, the suitable picture of atomic levels is the multi-term approximation of Sect.~7.6 of \citet{LL04} and the discussion in Sect.~13.4 therein.

For the synthesis of the D$_3$ line we use the code HAZEL \citep{2008ApJ...683..542A} which is applicable in the regime of our interest, namely in the limit of a constant-property slab. Given the height above the solar disk, magnetic field vector, and optical thickness of the slab obtained from the 
multi-ion non-LTE solution, we can calculate the line Stokes profiles and normalize the spectrum to the integrated absolute emissivity obtained from the unpolarized non-LTE solution (see Table.~\ref{table:models}). HAZEL uses the same five-term model atom as discussed above and takes properly into account the limb darkening effects in all the spectral lines.

At first, we consider the case of non-magnetic plasma ($B=0$\,G) and we calculate the dependence of the total (VL and D$_3$) integrated intensity and fractional polarization for different models. The results can be found in Figure~\ref{fig:iqb0}. We see a significant depolarization effect due to the D$_3$ line that is most apparent at intermediate temperatures around 30,000 K. This is closely related to the fact that at these temperatures the D$_3$ line is extremely bright, hence both the intensity and fractional polarization signal are dominated by the line instead of the VL (see Table~\ref{table:models}). Since fractional polarization of the line is always smaller than polarization of the VL, the presence of the line always leads to depolarization of the total signal (i.e. lowering of the fractional polarization). 
We note that fractional polarization of the line, $E_Q({\rm D}_3)/E_I({\rm D}_3)$ is practically insensitive to the model in the range of parameters of Table~\ref{table:models} and it only depends on the height above the solar surface and on the local magnetic field vector.
In the following section, we discuss the magnetic sensitivity of the total VL+D$_3$ signal.

\section{Magnetic Field and the  Hanle Effect}

\subsection{Flux-rope Magnetic Field in CMEs}

Measurements of magnetic fields in solar prominences are very difficult (see e.g. review by \cite{Lopez2015})
and usually require long integration times in order to detect weak polarization signals. 
To our knowledge, the magnetic field inside erupting prominences has never been measured during the eruption 
in the expansion phase across the intermediate corona (what will be observed by Metis).
However, in a recent paper by \cite{Fan+2018} the authors are simulating the 
appearance of an erupting prominence with the COSMO coronagraph, with the 
aim to demonstrate the feasibility of magnetic field determination from 
circular polarization measurements of the \ion{Fe}{13} line undergoing Zeeman 
effect. This paper provides the $B_{\rm LOS}$ 
(i.e. the LOS averaged magnetic field) inside an erupting prominence at 
different times during the eruption. From their Figs. 3, 
11, and 15, in the early eruption phase (below 1.4 solar radii from the disk center) one can 
assume maximum of 5-10 G at the center of the flux rope. 
For larger altitudes, we have to make some empirical considerations. Assuming
that the flux rope is expanding self-similarly, and using a well known empirical relationship 
between the radial speed $v_{\rm rad}$ and the expansion speed $v_{\rm exp}$ of a CME,
$v_{\rm rad} = 0.88 v_{\rm exp}$  \citep{DalLago+2003}, basically one can assume that
$h_{\rm CME} = 0.88 r_{\rm CME}$ and the same relationship holding between the altitude of 
the CME and the radius of the flux-rope. This means that the 
cross-sectional area $A_{\rm FR}$ of the flux rope is going with the CME 
altitude $h_{\rm CME}$ like
\begin{equation}
A_{\rm FR} = \pi r_{\rm CME}^2 = \pi (h_{\rm CME}/0.88)^2  \propto h_{\rm CME}^2 \, .
\end{equation}

Now, if we assume the magnetic flux conservation during the expansion, we 
can write that $B(0) A_{\rm FR}(0) = B(h) A_{\rm FR}(h)$. Therefore, as an order of magnitude 
estimate, if for instance $B$(0) = 10 G when $h_{\rm CME} = 1.2 R_{\rm sun}$ (Fig. 11 
from \cite{Fan+2018}) then for a CME at 3 $R_{\rm sun}$ 
the magnetic field is $B(h) = B(0) (1.2/3)^2 = 1.6$ G. One can use 
the same empirical relationship to rescale the field at higher/lower 
altitudes if required.

\subsection{Magnetic Sensitivity and the Simulated Hanle Diagrams}

\begin{figure} 
\begin{center}
\includegraphics[width=15.0cm]{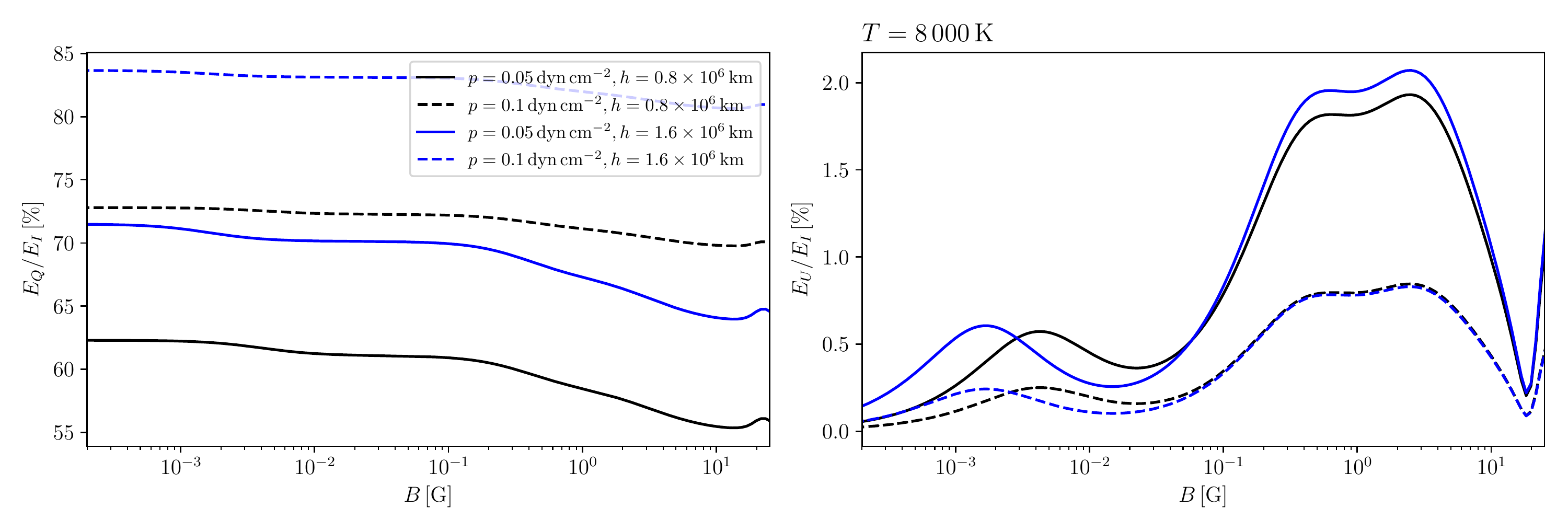}  \\
\includegraphics[width=15.0cm]{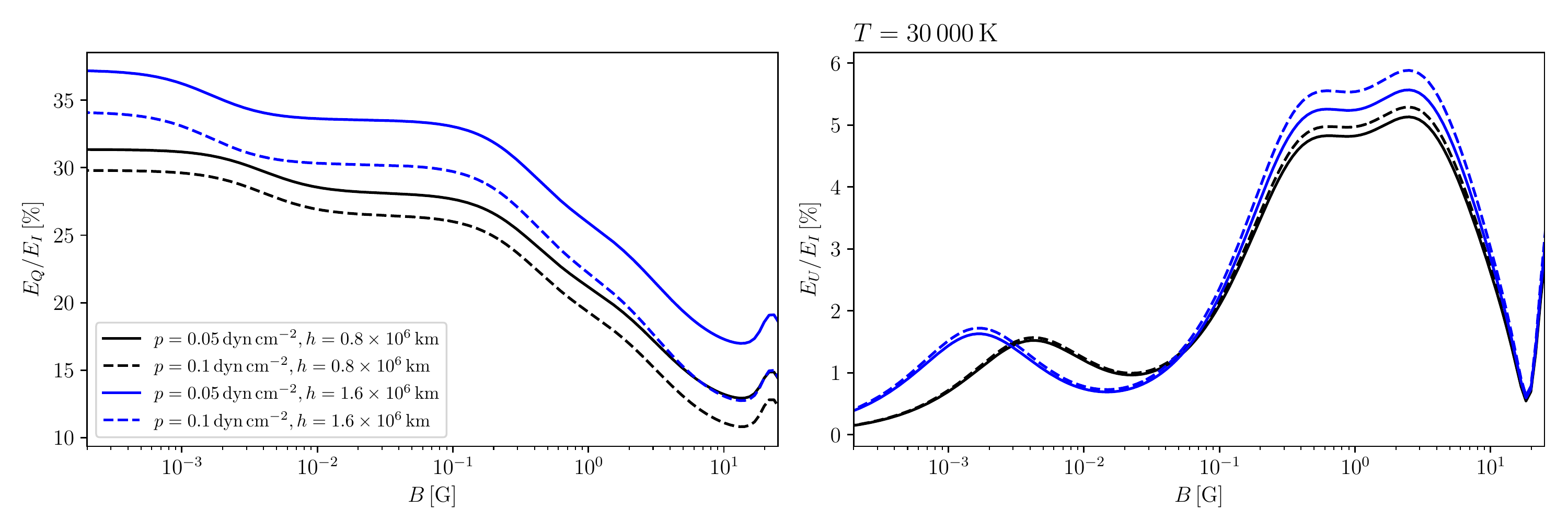} 
\includegraphics[width=15.0cm]{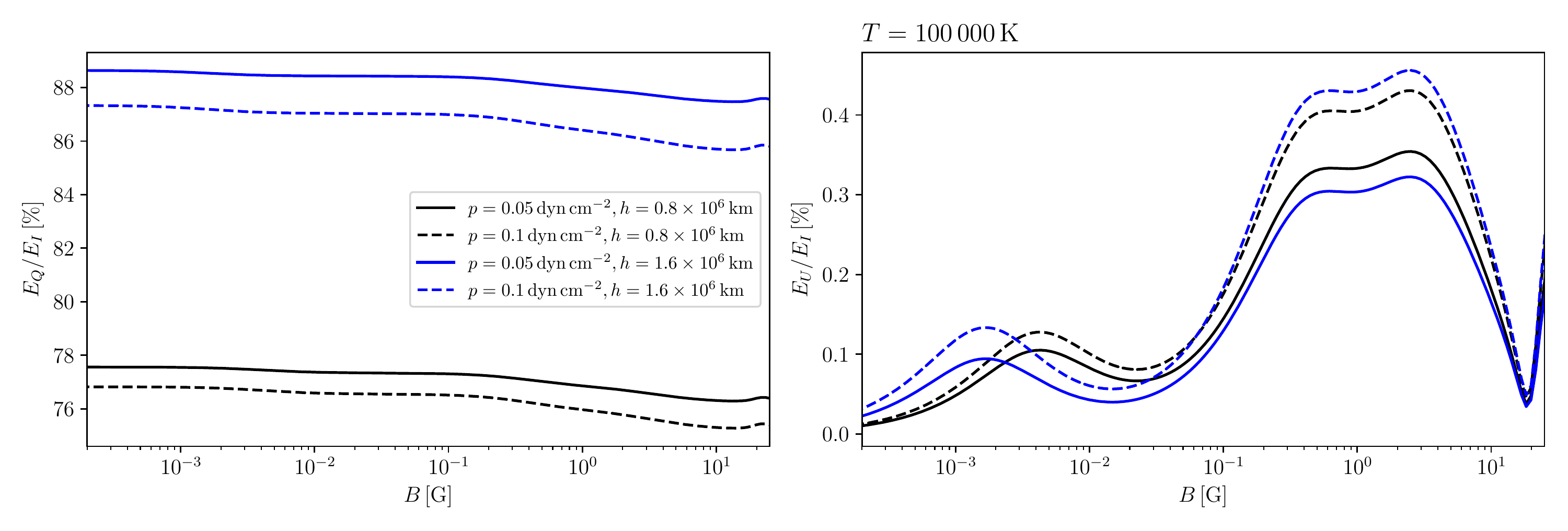}  \\
\end{center}
\caption{Total (i.e., VL and D$_3$) fractional polarization signals $E_Q/E_I$ (left panels) and $E_U/E_I$ (right panels) for different models as a function of magnetic field intensity. From top to bottom, the panels show models with temperatures 8,000, 30,000, and 100,000 K. The magnetic field is oriented  along the LOS ($\chi_B=0^\circ$).}
\label{fig:iqb}
\end{figure}

\begin{figure} 
\begin{center}
\includegraphics[width=5.5cm]{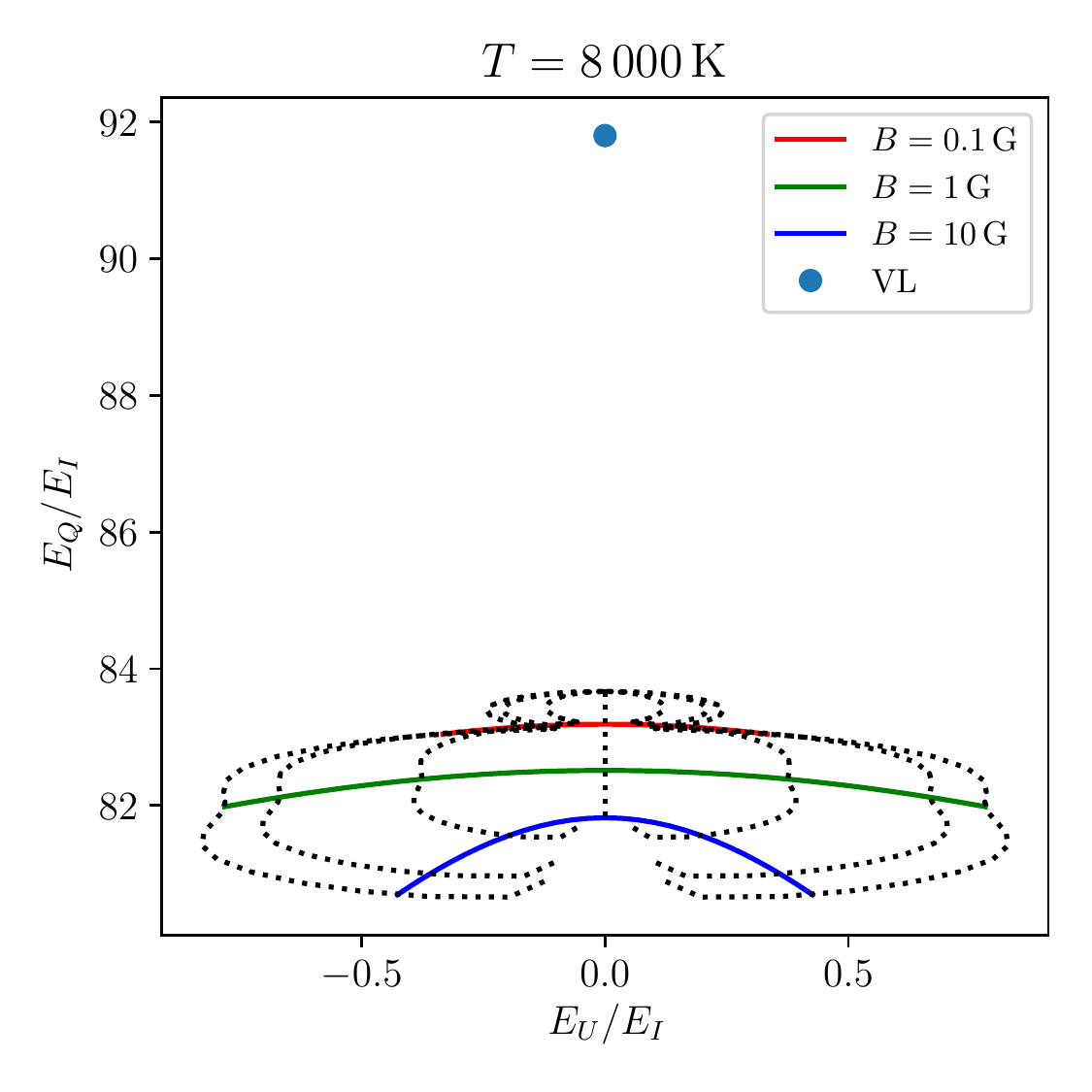}
\includegraphics[width=5.5cm]{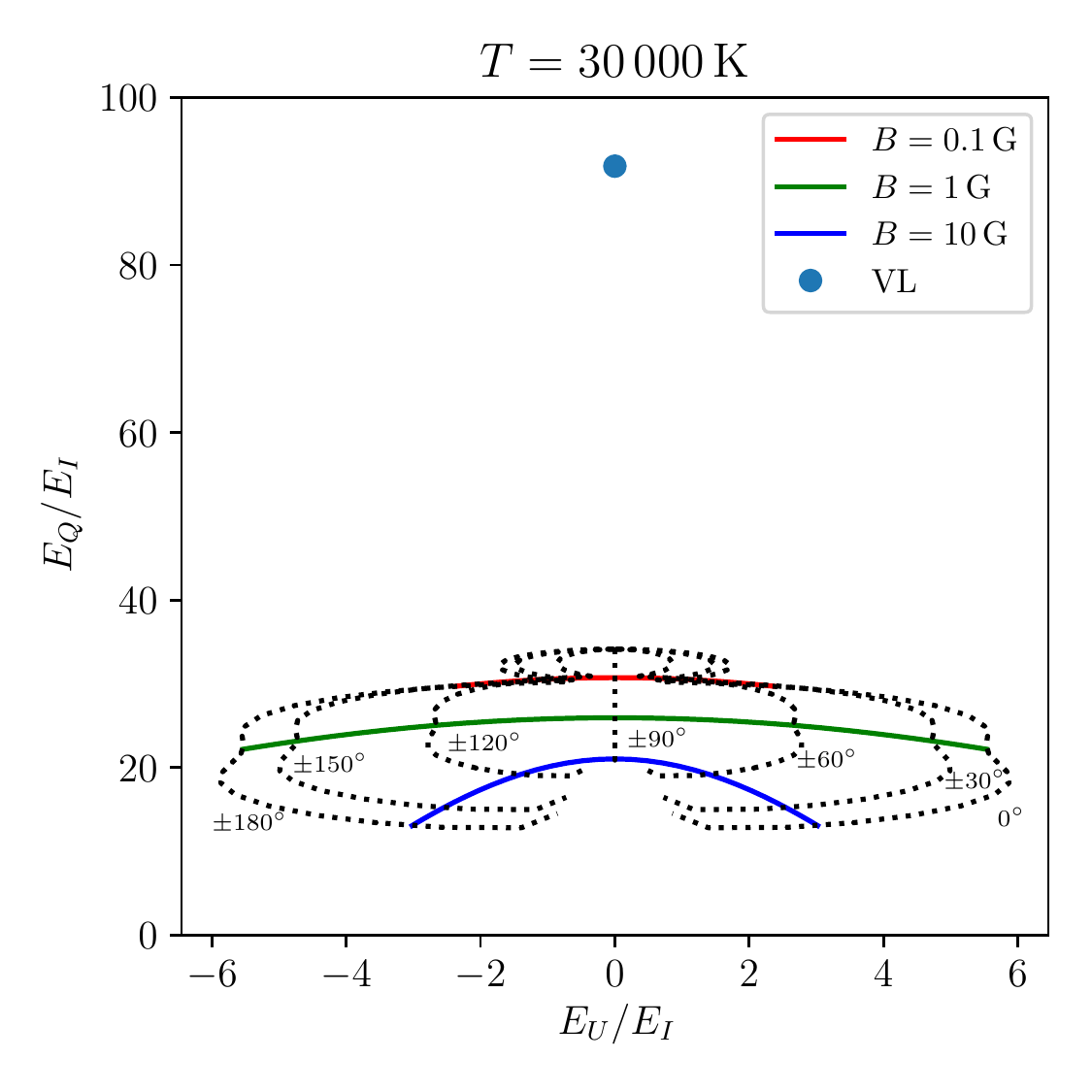}
\includegraphics[width=5.5cm]{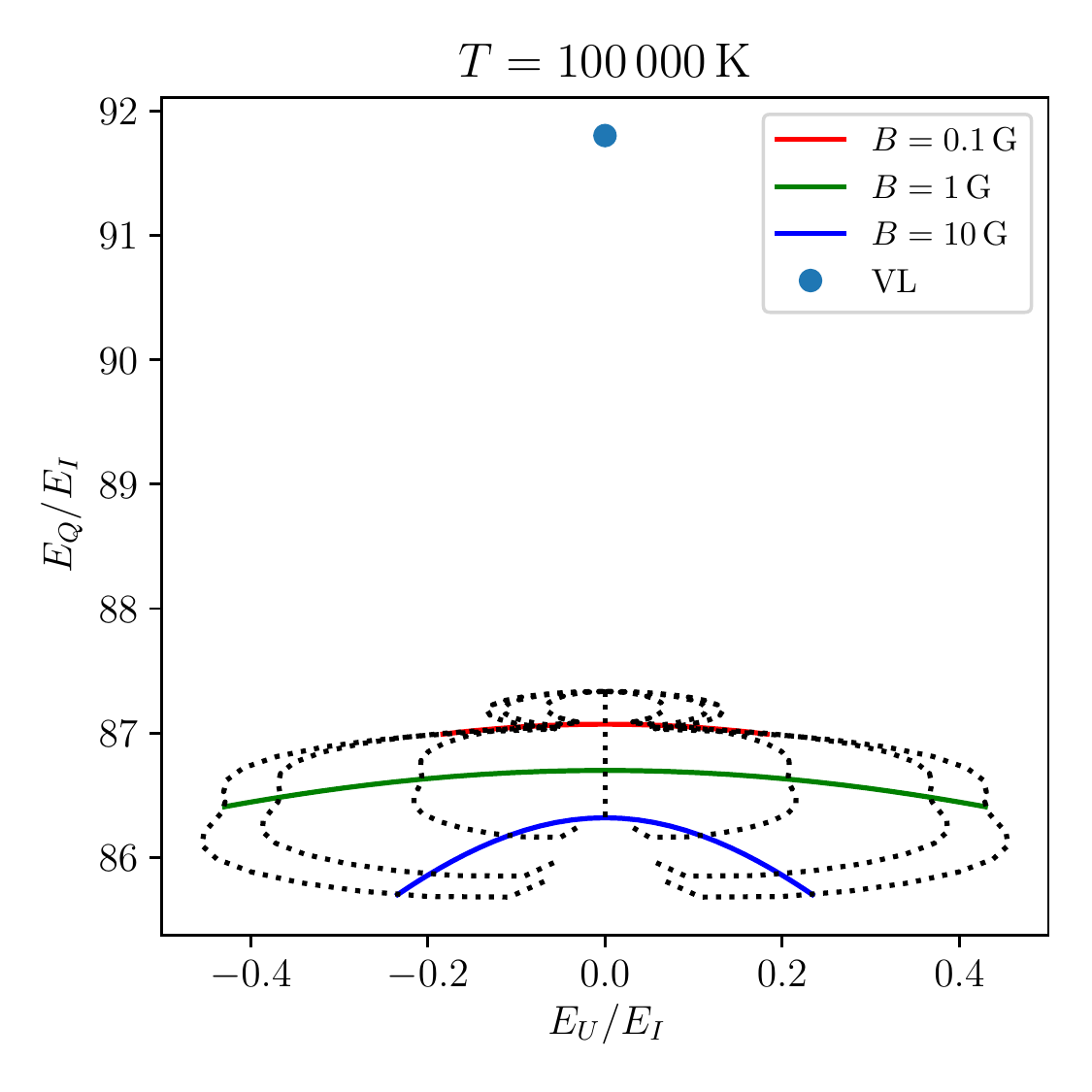}
\end{center}
\caption{Polarization (or Hanle) diagrams of the VL+D$_3$ signal for three different plasma temperatures, height $h=1.6\times 10^6$\,km, and $p=0.1\,{\rm dyn\,cm^{-2}}$.
The solid lines correspond to the magnetic field of a fixed strength and changing azimuth $\chi_B$ (indicated in the central panel). The dotted lines show the signals for fixed magnetic field azimuth and varying strength.
The signal of VL, i.e., neglecting the D$_3$ contribution, is shown by the blue dot on the top of the panels.}
\label{fig:hd}
\end{figure}

While the VL signal is insensitive to the magnetic field, the linear polarization of D$_3$ can be modified by the magnetic field due to the Hanle and Zeeman signals. In Figure~\ref{fig:iqb} we show the dependence of the total fractional linear polarization signals of the VL+D$_3$ on the magnetic field strength. Even a very weak magnetic field ($B<1$\,G) causes partial relaxation of the quantum coherence in the $2p\,{}^3 {\rm P}$ and $2s\,{}^3{\rm S}$ 
levels leading to the lower-level Hanle effect, hence to a sensitivity of the polarization signal to sub-G fields
\citep{Bommier1980}.
At fields of the order of few G, the signal is highly sensitive to the upper-term Hanle effect in $3d\,{}^3 {\rm D}$. 
Above $\approx 6$\,G, the $J$-level crossings start to occur and the upper term enters the so-called incomplete Paschen-Back effect \citep[see,
e.g.,][]{Sahal1981}.
These effects are most obvious in the central panels of Figure~\ref{fig:iqb} where the D$_3$ line dominates the total signal while in the cooller or hotter plasmas the depolarization and rotation of the polarization vector are less obvious.

The Hanle diagrams shown in Figure~\ref{fig:hd} further demonstrate the sensitivity of the linear polarization signal to different magnetic field strengths and orientations. It follows that the main impact of the D$_3$ line is a lowering of the fractional polarization $E_Q/E_I$  
due to relatively high D$_3$ intensity at temperatures around 30 - 50,000 K. However, for certain combinations
of magnetic field strength and orientation, the Hanle effect can leave measurable signatures in the linear polarization signal. For instance, at $B=3$\,G and $\chi_B=0^\circ$, the polarization vector can be rotated by about $10^\circ$ with respect to the nearest solar limb, thus providing a measurable positive evidence for the presence of magnetic field. In other cases, depolarization of the Stokes $Q$ parameter still provides a valuable constraint on the thermal conditions of the CME plasma.
 
\section{Discussion and Conclusions}

In order to detect signatures of the D$_3$ line emission in the integrated signal from the
VL filter of Metis, we must estimate the relative intensities of the line and VL continuum
which is due to Thomson scattering on free electrons in the CME core. The core is usually well
recognized as a bright flux rope or a patchy prominence-like structure, see e.g. \cite{Heinzel+2016} 
or \cite{FloydLamy2019} (their Figure 1) so that we can neglect other VL contributions due to
a CME hallo (quiet-corona emission is normally subtracted from images). Our theoretical estimates
for a range of plausible CME-core models are given in Table~\ref{table:models} where we can see that the D$_3$ 
contribution is non-negligible and in the temperature range between 30,000 and 50,000 K it
significantly dominates over the VL one (we discuss this behaviour in Section 3). We know that
such temperatures are present inside CME cores, see e.g. \cite{Jejcic+2017}. At 30,000 K, the D$_3$
line is a factor 5-7 brighter compared to VL. At low temperatures VL signal dominates and at 10$^5$ K
the line contribution is less that 10 \%. 

Normally one cannot infer a dominance of the line emission over the VL continuum just
from the fact that the CME core is structured like a cool prominence (see Figure 1 in \cite{FloydLamy2019} and the caption).
This is because
the CME-core electron densities will produce a similar pattern in VL due to Thomson scattering.
However, we can disentangle between these two
contributions using linear polarization measurements. Looking at Figure ~\ref{fig:iqb0} (right panel), 
we see that for temperatures
between say 20,000 and 80,000 K the polarization degree $E_Q/E_I$ is significantly lower compared
to a constant 80 - 90 \% polarization which is only due to Thomson scattering. This corresponds to an
intensity enhancement (peak) on the left panel of  Figure~\ref{fig:iqb0} which is due to D$_3$ emission.
Our models can thus, at least qualitatively, explain the behaviour of LASCO-C2 observations
analyzed by \cite{FloydLamy2019}. LASCO-C2 orange filter is even wider (540 - 640 nm) than the Metis VL filter, 
and also contains the D$_3$ line (in its central part). At least in two cases
analyzed by these authors, the detected polarization is surprisingly low compared to expected
high-degree polarization due to Thomson scattering alone. The authors thus conclude that this might be
due to a presence of the D$_3$ line and they think about possible depolarization due to Hanle effect. 
But since we know the brightness of the D$_3$ line in our models, we see that the line 
is sometimes very bright and therefore total $E_Q$ over total
$E_I$ is low even in a non-magnetic case. It is interesting to note that a low polarization was also
detected in a CME by \cite{Mierla+2011} using the wide-band filter of SECCHI/COR1 coronagraph which
contains the hydrogen H$\alpha$ line in its center. Note that this line is typically much brighter
in cool prominence structures as compared to D$_3$.

At this point we should mention that both Metis as well as LASCO-C2 filters contain \ion{Na}{1} doublet 
red-ward of the D$_3$ line at 589.16 nm and 589.76 nm (the rest wavelength). These lines are
also in emission in prominence-like structures and thus may contaminate the total VL signal.
However, as thoroughly discussed in \cite{Jejcic+2018}, their typical brightness is much lower
compared to D$_3$ line and thus we neglect them in our present analysis. 

Based on our estimation of the magnetic-field strength in CMEs, we studied the Hanle effect on
D$_3$ line. Concerning the orientation of the magnetic-field vector in a CME core, we can at least see
from numerous observations
that the expanding flux-rope top is more or less parallel to the solar limb and this is very
favorable for the field detection via the Hanle effect in the D$_3$ line. Our results are presented
and briefly discussed in Section 5. We show that under some conditions it will be possible to diagnose
CME magnetic field using the VL channel of Metis. 
At low temperatures (our models with 8,000 K) and for B around 10 G, the Hanle depolarization is only by about
2 \% (similar to quiescent prominences), at 10$^5$ K it is even lower.
But for intermediate temperatures the depolarization amounts to 10 \% or more - see polarization diagrams 
in Figure~\ref{fig:hd}. At 30,000 K we see a significant lowering of polarization degree just due to D$_3$ brightness
plus about 10 \% lowering due to 10 G magnetic field. The polarization capabilities of Metis are described
in \cite{Antonucci+2019} and \cite{Fineschi+2020} and in a following paper we plan to perform a more detailed 
analysis of Metis VL-channel response to 
magnetic-field and thermal structure of CMEs. In parallel we are studying a diagnostics
potential of the Metis Lyman $\alpha$ channel which can provide an independent information on CME
thermal structure. Finally, D$_3$ line intensity (i.e. Stokes $I$) and broad-band VL polarization will be
detected by the ASPIICS coronagraph on board of ESA's Proba-3 formation-flight mission and thus a synergy
with Metis will be of great interest.

\acknowledgements
{Solar Orbiter is the ESA mission towards the Sun launched by NASA on February 10, 2020.  
Metis is its payload coronagraph developed by the Italian-German-Czech consortium, under the
leadership of Italian ASI. 
The authors acknowledge support from the grants No. 19-17102S and 19-20632S of the Czech
Funding Agency and from RVO:67985815 project of the Astronomical Institute of the Czech Academy
of Sciences. The Czech contribution to Metis was funded by ESA-PRODEX.
S.J. acknowledges the support from the Slovenian Research Agency No. P1-0188.
N.L. acknowledges support from the Science and Technology Facilities Council (STFC)
Consolidated Grant No. ST/ P000533/1.}

\facility{Solar Orbiter - Metis}

\end{document}